\documentclass[epj,final]{svjour}
\usepackage{graphics}
\usepackage{amssymb}
\begin{document}
\title{Magnetic moment of an electron gas on the surface of constant negative curvature}
\author{D.V.~Bulaev \and V.A.~Margulis
}                     
\mail{bulaevdv@mrsu.ru}
\institute{Mordovian State University, Saransk, 430000 Russia}
\date{Received: date / Revised version: date}
%
\abstract{The magnetic moment of an electron gas on the surface of constant negative curvature is
investigated. It is shown that the surface curvature leads to the appearance of the region of the
monotonic dependence $M(B)$ at low magnetic fields. At high magnetic fields, the dependence of
the magnetic moment on a magnetic field is the oscillating one. The effect of the surface
curvature is to increase the region of the monotonic dependence of the magnetic moment and to
break the periodicity of oscillations of the magnetic moment as a function of an inverse magnetic
field.  
} 
\maketitle
\section{Introduction}
\label{intro}
The two-dimensional electron gas (2DEG) in quantized magnetic fields has attracted a lot of
attention in recent years. The increasing interest to the 2DEG is due to its unique properties:
periodic oscillations of the magnetic moment as a function of an inverse magnetic field (the de
Haas --- van Alphen effect), oscillations of the longitudinal magnetoresistance (the Shubnikov
--- de Haas effect), and quantization of the Hall conductivity (the integer and fractional
quantum Hall effects). Transport and magnetic measurements are powerful methods for studying the
electron energy spectrum. Note that transport measurements yield information about the localized
states near the Fermi energy, but provide very little direct information about the total density
of states of the 2DEG. In contrast, measurements of magnetization are a powerful tool for
exploring the total density of states of a degenerate Fermi gas. Recent advances in magnetometry
provide new insights into the electron density of states of the 2DEG \cite{HKM}.

The theory of the two-dimensional de Haas --- van Alphen effect was started by Peierls
\cite{Peierls}. In the zero-temp\-er\-at\-ure limit he obtained that the magnetization in the
ideal 2DEG has sharp, saw-tooth oscillations with a constant amplitude. Recent theoretical
investigations show that in the two-dimensional case the dependence of the oscillation amplitude
on a magnetic field and temperature differs from the Lifshitz --- Kosevich formula for bulk
conductors \cite{GV}, and this is confirmed by experiments \cite{HKM}.

The magnetic properties of electrons on curved surfaces have attracted a substantial interest in
recent years. This is due to the fact that these structures have unusual properties of the
magnetic moment as a function of a magnetic field, temperature, and geometric parameters
\cite{KVS,A59,BGM-00,MC,GMSH}.

In this paper we study the magnetic moment of electrons on the surface of constant negative
curvature (the Lobachevsky plane). Although the Lobachevsky plane has not been experimentally
realized yet, the problem of the physics on the Lobachevsky plane has a deep relation with some
interesting problems, like the occurrence of the chaos in the surface of negative curvature
\cite{Grosche2}, the Berry phase \cite{Albeverio}, and point perturbations \cite{Bruning} on the
Lobachevsky plane. In recent years, the quantum Hall effect on the Lobachevsky plane is a subject
of current interest \cite{Iengo,Avron,Pnueli,BGM-03}.

\section{Electron states and magnetic moment}
\label{sec:1}
We consider noninteracting electrons confined to the surface of constant negative curvature (the
Lo\-ba\-chev\-sky plane) in a magnetic field $\vec{B}$. We choose the Landau gauge for the vector
potential ($\textbf{A}=(Ba^2y^{-1},0)$) and employ the Poincar\'e realization in which the
Lobachevsky plane $L$ is identified with the upper complex halfplane $L=\{z=x+iy\in \mathbb{C}:\
y>0\}$ endowed with the metric
\[
ds^2=\frac{a^2}{y^2}(dx^2+dy^2),
\]
where $a$ is the radius of curvature. The Hamiltonian of the system can be written as
\cite{BGM-03}
\begin{equation}
\label{eq:Hamiltonian}
H=\frac{\hbar^2}{2m^*a^2}\left[-y^2\left(\partial_x^2+
\partial_y^2\right)+2iby\partial_x+b^2-\frac14\right],
\end{equation}
where  $m^*$ is the effective electron mass, $b=eBa^2/\hbar c$, and the last term in
Eq.~(\ref{eq:Hamiltonian})  is the surface potential which arises from the surface curvature
\cite{Lan}. The spectrum of $H$ consists of two parts \cite{Comtet}: a discrete spectrum in the
interval $(0,\hbar^2b^2/2m^*a^2)$ consisting of a finite number of Landau levels
\begin{equation}
\label{eq:Energy}
E_n=\hbar\omega_c\left(n+\frac12\right)-\frac{\hbar^2}{2m^*a^2} \left(n+\frac12\right)^2,\ 0\le
n<|b|-\frac12
\end{equation}
and a continuous spectrum in the interval $[\hbar^2b^2/2m^*a^2,\infty)$
$$
E(\nu)=\frac{\hbar^2}{2m^*a^2}\left(b^2+\nu^2\right),\ 0\le\nu<\infty.
$$

The electron density of states $n(E)$ per unit area is defined by the following expression:
\[
n(E)=\frac{1}{\pi S}\int \mathrm{Im} G(\vec{r},\vec{r};E+i0)d\vec{r},
\]
where $S$ is the area of the surface and $G(\vec{r},\vec{r'};E)$ is the Green's function of the
Hamiltonian. The density of states of electrons on the Lobachevsky plane is given by
\cite{BGM-03}
\begin{eqnarray}\nonumber
&&n(E)=\frac{1}{2\pi a^2}\sum_{0\le n<|b|-1/2}
\left(|b|-n-\frac12\right)\delta(E-E_n)\\
\nonumber&&+\frac{m^*}{2\pi\hbar^2}
\Theta\left(E-\frac{\hbar^2b^2}{2m^*a^2}\right)\\
\label{eq:DOS}&&\times\frac{\sinh 2\pi \sqrt{2m^*a^2E/\hbar^2-b^2}} {\cosh 2\pi\sqrt{2m^*a^2E/\hbar^2-b^2}+\cos
2\pi b},
\end{eqnarray}
where $\Theta(x)$ is the Heaviside step function. The first term in Eq.~(\ref{eq:DOS})
corresponds to the discrete spectrum and the second term corresponds to the continuous one and
coincides with the expression given in Ref.~\cite{Comtet}.

The magnetic moment of a thermodynamic system with a fixed chemical potential is given by
\begin{equation}
\label{eq:M}
\textbf{M}=-\left.\left(\frac{\partial\Omega}{\partial\textbf{B}}\right)\right|_{T,S,\mu},
\end{equation}
where
\begin{equation}\label{eq:Omega}
\Omega(T,\mu)=-TS\int_{-\infty}^\infty n(E)
\ln\left\{1+\exp\left[\left(\mu-E\right)/T\right]\right\}dE
\end{equation}
is the thermodynamic potential.

First, we consider the case of zero temperature. We substitute the formula for the electron
density of states (Eq.~(\ref{eq:DOS})) into Eq.~(\ref{eq:Omega}) and apply the Poisson summation
formula for the magnetic moment. When both the discrete spectrum and the continuous one are below
the Fermi energy, i.e, when $\varepsilon_F\ge b^2$ ($\varepsilon_F=2m^*a^2E_F/\hbar^2$), the
magnetic moment of an 2DEG on the Lobachevsky plane is given by
\begin{eqnarray}\nonumber
&&\left.\frac{M(T=0)}{\mu_B}\right|_{\varepsilon_F\ge b^2}=\frac{m_e}{m^*}\frac{S}{2\pi
a^2}\left\{-\frac b6\right.\\
\nonumber&&-\frac{3}{2\pi^3}\sum_{k=1}^\infty\frac{(-1)^k}{k^3}\exp[-2\pi
k\sqrt{\varepsilon_F-b^2}]\sin 2\pi kb\\
\nonumber &&+\frac{2}{\pi}\sqrt{\varepsilon_F-b^2}\sum_{k=1}^\infty\frac{(-1)^k}{k}\exp[-2\pi
k\sqrt{\varepsilon_F-b^2}]\\
\nonumber&&\times\left(b\cos 2\pi kb-\sqrt{\varepsilon_F-b^2} \sin 2\pi kb\right)\\
\nonumber&&+\frac{1}{\pi^2}\sum_{k=1}^\infty\frac{(-1)^k}{k^2}\exp[-2\pi
k\sqrt{\varepsilon_F-b^2}]\\
\label{eq:M1}&&\left.\times\left(b\cos 2\pi kb-3\sqrt{\varepsilon_F-b^2} \sin 2\pi
kb\right)\right\}.
\end{eqnarray}
As can be seen from Eq.~(\ref{eq:M1}), in the region $\varepsilon_F\ge b^2$, $M(B)$ is the
monotonic dependence, since the exponents in the sums are negligible.

For $\varepsilon_F<b^2$ (in this case only the discrete spectrum is below the Fermi energy), the
magnetic moment is given by
\begin{eqnarray}\nonumber
&&\left.\frac{M(T=0)}{\mu_B}\right|_{\varepsilon_F<b^2} =\frac{m_e}{m^*}\frac{S}{2\pi a^2}
\left\{-\frac b6\right.\\
\nonumber&&-\frac{3}{2\pi^3}\sum_{k=1}^\infty\frac{(-1)^k}{k^3}
\sin 2\pi k(b-\sqrt{b^2-\varepsilon_F})\\
\nonumber &&+\frac{1}{\pi^2}(b-3\sqrt{b^2-\varepsilon_F})
\sum_{k=1}^\infty\frac{(-1)^k}{k^2}\cos 2\pi k(b-\sqrt{b^2-\varepsilon_F})\\
\nonumber &&-\frac{2}{\pi}\sqrt{b^2-\varepsilon_F}(b-\sqrt{b^2-\varepsilon_F})\\
\label{eq:M2}&&\times\left.\sum_{k=1}^\infty\frac{(-1)^k}{k}\sin 2\pi k(b-\sqrt{b^2-\varepsilon_F})\right\}.
\end{eqnarray}
As can be seen from Eq.~(\ref{eq:M2}), the dependence $M(B)$ has the monotonic part, which is a
linear function of a magnetic field, and the three sums lead to saw-tooth oscillations.

\begin{figure}
\resizebox{0.9\columnwidth}{!}{%
  \includegraphics{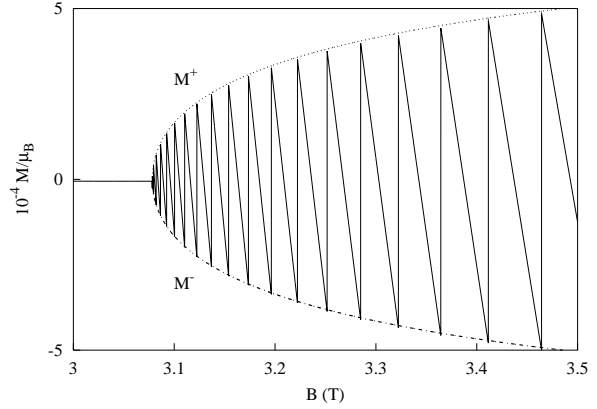}
}

\caption{The magnetic moment of a 2DEG on the Lobachevsky plane as a function of a magnetic
field; $a=10^{-5}\;$cm, $S=3\times10^{-9}\;$cm$^2$, $E_F = 2\times10^{-13}\;$erg.}
\label{fig:1}
\end{figure}

In Fig.~\ref{fig:1} we show the dependence $M(B)$. The dotted and dot-dashed lines show the
envelope of maxima and minima of the magnetic moment respectively.

The monotonic dependence $M(B)$ corresponds to the case when both the discrete spectrum and the
continuous one are below the Fermi energy. The bottom of the continuous spectrum crosses the
Fermi level with increasing a magnetic field, therefore, there is only the discrete spectrum
below the Fermi energy. In this case, the monotonic dependence of the magnetic moment on a
magnetic field is replaced by the oscillating one. As can be seen from Fig.~\ref{fig:1}, for
$\varepsilon_F<b^2$, the monotonic part of the magnetic moment is much less than the amplitude of
saw-tooth oscillations. The jumps in the magnetic moment arise from the crossings of the Fermi
level by the electron levels.

Let us consider the monotonic dependence  $M(B)$. As can be seen from Eq.~(\ref{eq:M1}), the
exponents in the sums are negligible for $\varepsilon_F\gg b^2$. Neglecting these sums, we have
that the dependence of the magnetic moment on a magnetic field is almost linear
\begin{equation}\label{eq:Mlinear}
\left.\frac{M(T=0)}{\mu_B}\right|_{\varepsilon_F\gg b^2}\simeq \frac{m_e}{m^*}\frac{S}{2\pi
a^2}\left(-\frac b6\right).
\end{equation}

Now consider the magnetic moment oscillations in the region $\varepsilon_F<b^2$. From
Eq.~(\ref{eq:M2}) it is easy to find the envelope of the maxima and minima of the magnetic moment
$$
\frac{M^\pm}{\mu_B}=\frac{m_e}{m^*}\frac{S}{2\pi a^2}\sqrt{b^2-\varepsilon_F} \left(\pm
b-\frac12\mp\sqrt{b^2-\varepsilon_F}\right),
$$
where $M^+$ and $M^-$ are the envelopes of the maxima and minima of the magnetic moment
respectively. The difference between them is the oscillation amplitude:
$$
\frac{\Delta M}{\mu_B}=\frac{M^+-M^-}{\mu_B}= 2\frac{m_e}{m^*}\frac{S}{2\pi
a^2}\sqrt{b^2-\varepsilon_F} \left(b-\sqrt{b^2-\varepsilon_F}\right).
$$
As can be seen from this equation, the surface curvature decreases the amplitude of oscillations
of the magnetic moment. The oscillation amplitude tends to $E_F2m_eS/\pi\hbar^2$ with a magnetic
field. As shown in Fig.~\ref{fig:1}, oscillations of the magnetic moment as a function of a
magnetic field are not periodic. It is easy to find the distance between two neighboring jumps in
units of $1/B$:
$$
\Delta\left(\frac 1B\right)=\frac{|e|a^2}{\hbar
c}\frac{2\sqrt{b^2-\varepsilon_F}-1}{b\left(b^2+2b+1-2(b+1)\sqrt{b^2-\varepsilon_F}\right)}.
$$
As can be seen form this equation, the oscillations of the magnetic moment as a function of an
inverse magnetic field ($1/B$), in contrast with the case of the flat surface, are not periodic.
Note that in the limit of zero curvature the distance between two neighboring jumps is given by
$$
\Delta\left(\frac 1B\right)\mathop{\longrightarrow}\limits_{a\to\infty}\frac{\hbar|e|}{m^* c
E_F}.
$$

From an analytical study of the obtained formulas and from a numerical analysis we find that the
region of the monotonic dependence of the magnetic moment is increasing, the amplitude and the
distance between neighboring jumps of the magnetic moment as a function of a magnetic field are
decreasing with increasing surface curvature.

Let us now consider the effect of temperature on the magnetic moment of a 2DEG on the Lobachevsky
plane. Substituting Eq.~(\ref{eq:DOS}) into Eq.~(\ref{eq:Omega}), we get $M=M_1+M_2$, where
\begin{eqnarray*}
&&\frac{M_1}{\mu_B}=-\frac{m_e}{m^*}\frac{S}{2\pi a^2}\left\{ \sum_{0\le
n<b-1/2}\left(b-n-\frac12\right)(2n+1)f_0(E_n)\right.\\
&&\left.-\frac{2m^*a^2T}{\hbar^2}\sum_{0\le n<b-1/2} \ln\left\{1+\exp[(\mu-E_n)/T]\right\}
\right\}
\end{eqnarray*}
is the contribution of the discrete spectrum,
\begin{eqnarray*}
&&\frac{M_2}{\mu_B}=-\frac{m_e}{m^*}\frac{S}{2\pi a^2}\left\{b\int_0^\infty dt
\frac{\sinh2\pi\sqrt{t}}{\cosh2\pi\sqrt{t}+\cos2\pi
b}\right.\\
\nonumber&&\times f_0\left(\frac{\hbar^2}{2m^*a^2}(t+b^2)\right)\\
&&-\frac{2m^*a^2T}{\hbar^2}\pi\sin2\pi b \int_0^\infty dt
\frac{\sinh2\pi\sqrt{t}}{(\cosh2\pi\sqrt{t}+\cos2\pi b)^2}\times\\
&&\left.\times\ln\left\{1+\exp\left[\left(\mu-\frac{\hbar^2}{2m^*a^2}(t+b^2)\right)/T\right]\right\}
\right\}
\end{eqnarray*}
is the contribution of the continuous spectrum to the magnetic moment. The numerical study shows
that temperature results in smearing of the oscillations maxima and decreasing the oscillation
amplitude. The monotonic part of the magnetic moment is nearly independent of temperature.

\section{Conclusions}
\label{sec:2}
The magnetic moment of a 2DEG on the surface of constant negative curvature is investigated. It
is shown that when both the discrete spectrum and the continuous one are below the Fermi energy,
the dependence of the magnetic moment on a magnetic field is the monotonic one.  The bottom of
the continuous spectrum crosses the Fermi level with increasing a magnetic field. In this case,
the monotonic dependence of the magnetic moment on a magnetic field is replaced by the
oscillating one. The effect of the surface curvature is to increase the region of the monotonic
dependence of the magnetic moment and to decrease the amplitude and the distance between
neighboring jumps of the magnetic moment as a function of a magnetic field.

\begin{acknowledgement}
This work was supported by the INTAS Grant No.~00--257 and the Russian Ministry of Education
Grant No.~E02--3.4--370.  One of us (D.V.B.) was supported in part by the Russian Ministry of
Education Grant No.~E02--2.0--15, and the RFBR Grant No.~03-02-06006-mas.
\end{acknowledgement}


\begin{thebibliography}{}

\bibitem{HKM}
J.~Harris et al., Phys. Rev. Lett.
\textbf{86}, 4644 (2001)

\bibitem{Peierls}
R.~Peierls, Phys. Z \textbf{81}, 186 (1933)

\bibitem{GV}
D.~Grigoriev, I.D.~Vagner, Pis'ma v ZhETF \textbf{69}, 139 (1999) [JETP Lett. \textbf{69}, 156
(1999)]

\bibitem{KVS}
J.H.~Kim, I.D.~Vagner, B.~Sundaram, Phys. Rev. B \textbf{46}, 9501 (1992)

\bibitem{A59}
D.N.~Aristov, Phys. Rev. B \textbf{59}, 6368 (1999)

\bibitem{BGM-00}
D.V.~Bulaev, V.A.~Geyler, V.A.~Margulis, Phys. Rev. B \textbf{62}, 11517 (2000)

\bibitem{MC}
L.I.~Magarill, A.V.~Chaplik, Zh. Eksp. Teor. Fiz. \textbf{115}, 1478 (1999) [Sov. Phys. JETP
\textbf{88}, 815 (1999)]

\bibitem{GMSH}
V.A.~Geyler, V.A.~Margulis, A.V.~Shorokhov, Zh. Eksp. Teor. Fiz. \textbf{115}, 1450 (1999) [Sov.
Phys. JETP \textbf{88}, 800 (1999)]

\bibitem{Grosche2}
C.~Grosche, J. Phys. A \textbf{25}, 4573 (1992)

\bibitem{Albeverio}
S.A.~Albeverio, P.~Exner, V.A.~Geyler, Lett. Math. Phys. \textbf{55}, 9 (2001)

\bibitem{Bruning} J.~Br\"uning, V.A.~Geyler, Teor. Matem. Fiz. \textbf{115}, 368 (1999)
 [Theor. Math. Phys. \textbf{119}, 687 (1999)]

\bibitem{Iengo}
R.~Iengo, D.~Li, Nucl. Phys. B, \textbf{413}, 735 (1994)

\bibitem{Avron}   J.E.~Avron et al., Phys. Rev. Lett. \textbf{69}, 128 (1992)

\bibitem{Pnueli}  A.~Pnueli, Ann. Phys. \textbf{231}, 56 (1994)


\bibitem{BGM-03} D.V.~Bulaev, V.A.~Geyler, V.A.~Margulis, Phisica B (to be published)

\bibitem{Lan}      N.P.~Landsman, \textit{Mathematical topics between classical and quantum mechanics\/}
(Springer-Verlag, New York, 1998)

\bibitem{Comtet}   A.~Comtet,
                   Ann. Phys. \textbf{173}, 185 (1987)


\end{thebibliography}
\end{document}